\documentclass[useAMS,usenatbib]{mn2e}
\usepackage{graphicx}

\title[1.4~GHz polarimetric observations of the two fields imaged by
	the DASI experiment]
      {1.4~GHz polarimetric observations of the two fields imaged by
	the DASI experiment}
\author[G. Bernardi, et al.]
{G.~Bernardi$^{1}$\thanks{E-mail:
bernardi@iasfbo.inaf.it}, 
E. Carretti$^{1}$,
R.J. ~Sault$^{2}$, 
S.~Cortiglioni$^{1}$ and S.~Poppi$^{3}$\\
$^{1}$INAF--IASF Bologna, Via Gobetti 101, Bologna, I-40129, Italy\\
$^{2}$CSIRO--ATNF, P.O. Box 76, Epping, NSW 1710, Australia\\
$^{3}$INAF--IRA Bologna, Via Gobetti 101, Bologna, I-40129, Italy}
\begin{document}

\date{Accepted xx xx xx. Received yy yy yy; in original form zz zz zz}

\pagerange{\pageref{firstpage}--\pageref{lastpage}} \pubyear{2005}

\maketitle

\label{firstpage}

\begin{abstract}
We present results of polarization observations at 1.4~GHz of the
two fields imaged by the DASI experiment ($\alpha = 23^{\rm h}
30^{\rm m}$, $\delta = -55^{\circ}$ and $\alpha = 00^{\rm h} 30^{\rm m}$,
$\delta = -55^{\circ}$, respectively). Data were taken with the Australia
Telescope Compact Array with 3.4~arcmin resolution and $\sim
0.18$~mJy~beam$^{-1}$ sensitivity. The emission is dominated by point sources
and we do not find evidence for diffuse synchrotron radiation even after source
subtraction. This allows to estimate an upper limit of the diffuse
polarized emission. The extrapolation to 30~GHz suggests that the synchrotron
radiation is lower than the polarized signal measured by the DASI
experiment by at least 2 orders of magnitude. This further supports the conclusions
drawn by the DASI team itself about the negligible Galactic foreground
contamination in their data set, improving by a factor $\sim 5$ the upper limit
estimated by Leitch et al. (2005). 

The dominant point source emission allows us to estimate the
contamination of the CMB by extragalactic foregrounds. We computed the power
spectrum of their contribution and its extrapolation to 30~GHz provides a
framework where the CMB signal should dominate. However, our results do not
match the conclusions of the DASI team about the negligibility of point source
contamination, suggesting to take into account a source subtraction from the
DASI data.
\end{abstract}

\section{Introduction}
The key role played by the Cosmic Microwave Background Polarization (CMBP) in
modern cosmology is now well established. The CMB $E$--mode carries information
on the reionization (e.g. Zaldarriaga 1997) and, joined to the temperature
anisotropies, helps constraining the cosmological parameter estimation
(e.g. Zaldarriaga, Spergel \& Seljak 1997). The CMB $B$--mode looks interesting
for its connection with the inflationary era (e.g. Kosowsky 1999) and its 
capability to measure the amount of primordial gravitational waves.

On the experimental hand, the search for the CMBP is under way. Detections
have been claimed by the DASI (Leitch et al. 2005), CAPMAP (Barkats et al.
2004), CBI (Readhead et al. 2004) and BOOMERanG (Montroy et al. 2005) teams.
However, we are far from a complete characterization of the $E$--mode power
spectrum whereas the $B$--mode detection is still missing. 

The tiny level of the CMB polarized components makes the contamination by
foreground astrophysical sources even more severe than for the anisotropy
temperature term.

At  frequencies lower than 100~GHz, the foreground contribution is expected to
be dominated by both the Galactic synchrotron emission and extragalactic
radiosources. 

Several theoretical works (Mesa et al. 2002, Tucci et al. 2004, de Zotti et
al. 2005) studied the contamination due to point source emission on the
microwave background using data coming from low frequencies surveys. In
particular Mesa et al. (2002) and Tucci et al. (2004) achieve similar results
pointing out that the contamination to the CMB $E$--mode is not serious in the
70--100~GHz frequency range. However, both of them find that extragalactic
sources could be a considerable contaminant up to 44~GHz. In addition,
observational efforts are underway in order to establish the properties of the
radio sources directly at high frequency. The results from 18~GHz observations
of the Kuhr sample (Ricci et al. 2004a) and preliminary results from a
18~GHz survey by Ricci et al. (2004b) show a substantial agreement with the
predictions by Mesa et al. (2002). Given this, the estimate of the $E$--mode
contamination by point sources is still open for frequencies up to 44~GHz.  

About the Galactic synchrotron emission, template maps (e.g. Giardino et al.
2002, Bernardi et al. 2004) suggest that the CMBP $E$--mode signal should be
higher than the synchrotron emission on large angular scales and for frequencies
around 90~GHz.

On degree and sub--degree scales, polarization observations of the diffuse
synchrotron radiation in 
selected low emission regions have just recently begun with the aim of
estimating the possible contamination on the CMBP. The field imaged by the
BOOMERanG experiment was observed with both the Australia Telescope Compact
Array (ATCA) at 1.4~GHz (Bernardi et al. 2003, hereafter B03, Carretti et al.
2005a) and the Parkes telescope at 2.3~GHz (Carretti et al. 2005b) providing the
first characterization of the diffuse radiation in a low emission area. They
find that the Galactic signal should not prevent the detection of the CMB
$E$--mode in that area at frequencies $\ge 30$~GHz. A similar analysis has been
conducted in another area in the Northern sky providing similar conclusions
(Carretti et al. 2006). Although these results can be used as indicators of
conditions in low emission regions, every area has its own specific features and
requires dedicated observations, especially if representing a target for CMBP
experiments.

Among the sky regions imaged by CMB experiments, direct observations of the
synchrotron emission only exist for BOOMERanG area (B03) to date. Barkats et al.
(2005) 
assessed that the CAPMAP field is not contamined at 90~GHz since they checked
that the total intensity synchrotron emission in their area is below the
detected polarized signal. On the other hand, DASI and CBI operated at $\sim
30$~GHz, where the synchrotron emission can still play a relevant role. In
particular, the DASI team itself pointed out that no polarization observations
are available for the two sky patches they observed (Kovac et al. 2002).

In this paper we present results of the first deep polarization observations of
the two fields imaged by the DASI experiment at a frequency of 1.4~GHz, where
the synchrotron emission is dominant.

The paper is organized as follows: in Section~\ref{dasi_observations} we
describe the observations while in Section~\ref{CMBsec} we present the power
spectrum analysis. Finally, in Section~\ref{discussion_CMB} we discuss the
results in the framework of CMBP measurements.

\section{Observations and results}\label{dasi_observations}

\begin{figure*}
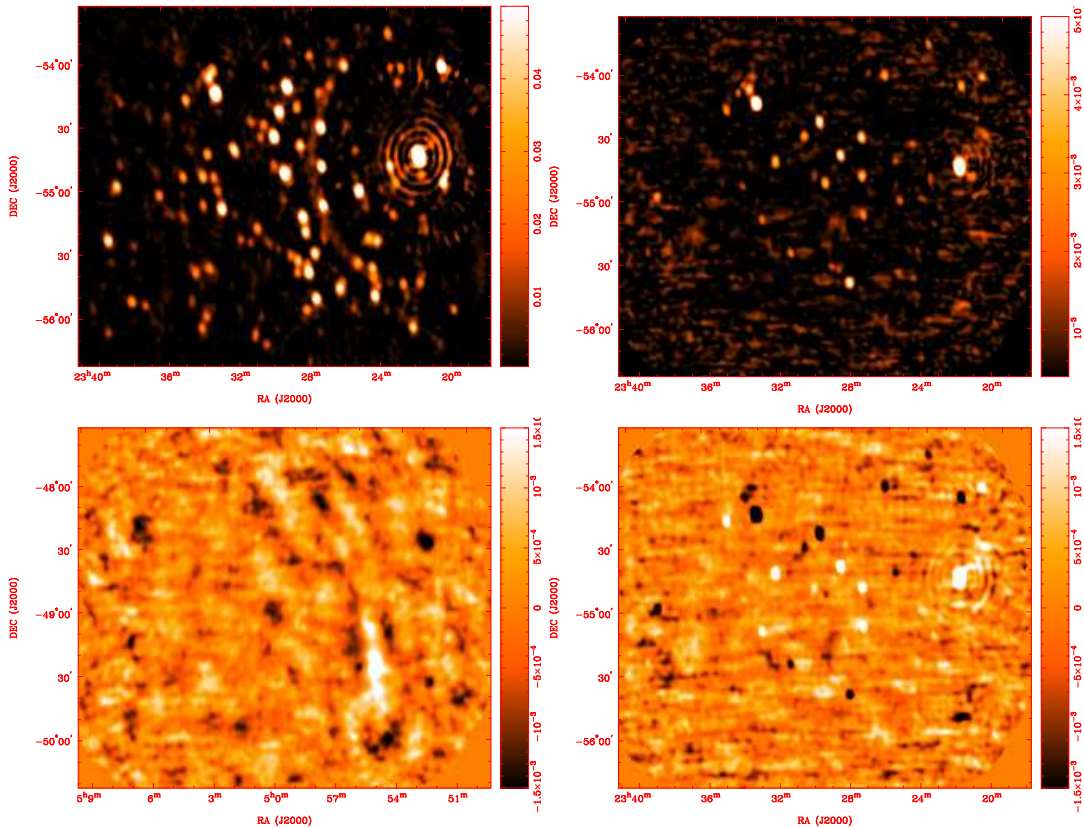

  \includegraphics[angle=-90, width=0.4\hsize]{dasi1_i.ps}
  \includegraphics[angle=-90, width=0.4\hsize]{dasi1_pol.ps}
  \includegraphics[angle=-90, width=0.4\hsize]{boom_u.ps}
  \includegraphics[angle=-90, width=0.4\hsize]{dasi1_u_051031.ps}
\caption{DASI field 1: total intensity $I$ (top-left), polarized intensity $I^p
= \sqrt{Q^2+U^2}$ (top-right). The $U$ image (bottom-right) is also reported for
comparison with the diffuse emission (point sources subtracted) detected in
B03 (bottom-left).\label{dasi1_images}}
\end{figure*}
\begin{figure*}
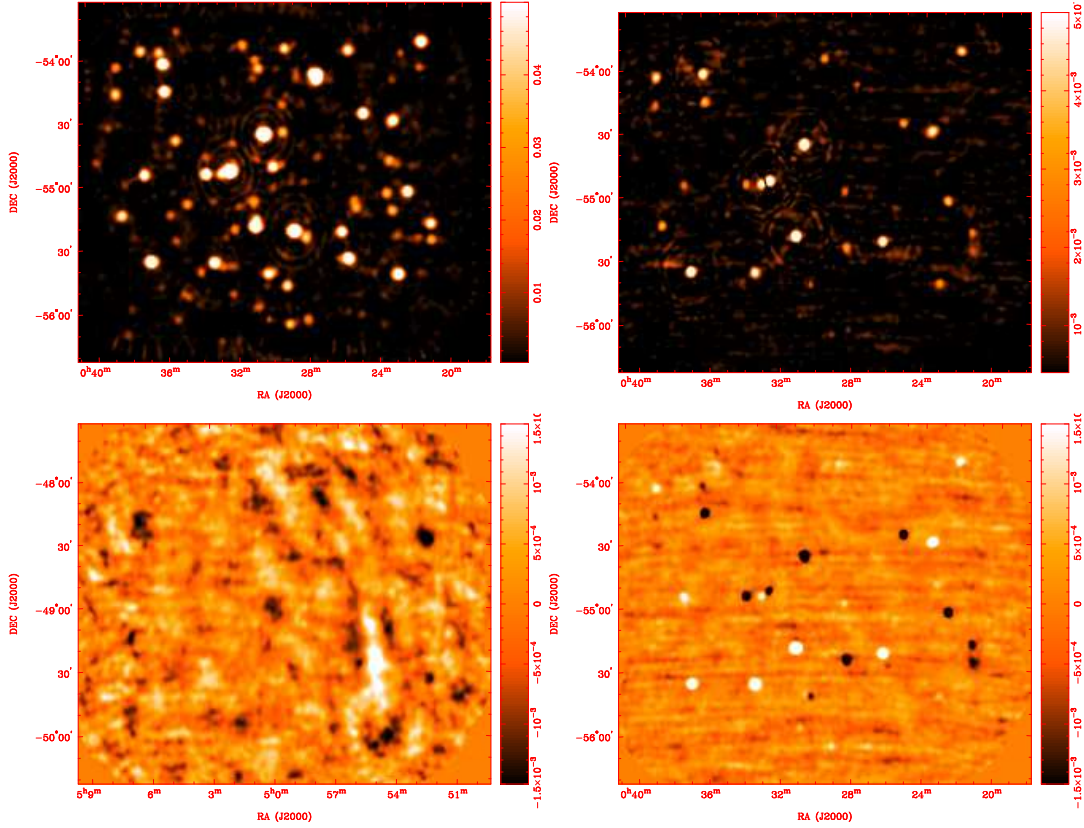

  \includegraphics[angle=-90, width=0.4\hsize]{dasi2_i.ps}
  \includegraphics[angle=-90, width=0.4\hsize]{dasi2_pol.ps}
  \includegraphics[angle=-90, width=0.4\hsize]{boom_u.ps}
  \includegraphics[angle=-90, width=0.4\hsize]{dasi2_u_051031.ps}
\caption{The same as Figure~\ref{dasi1_images} but for the field
	2.\label{dasi2_images}} 
\end{figure*}

Observations of the two DASI fields (hereafter named field 1 and field 2, see
Table~\ref{obsTab} for their coordinates) were performed in July 2003
and September 
2004 with the ATCA. The EW214 configuration was used to sample the spatial
frequencies between 3 and 30~arcmin. Each field has a $3^\circ \times
3^\circ$ size and the same amount of time ($\sim 72$~h) was spent in each
of them. A sensitivity of $\sim 0.18$~mJy/beam has been achieved in
both fields in the inner $2^\circ \times 2^\circ$ square, corresponding to $\sim
3.2$~mK. Both of them were observed with the same system configuration whose
characteristics are reported in Table~\ref{obsTab}.
\begin{table}
 \centering
  \caption{Main characteristics of the observations.}
  \begin{tabular}{@{}lr@{}}
  \hline
  Central Frequency &  1380~MHz \\
  Effective Bandwidth & $205$~MHz \\
  Field 1 (J2000) & $\alpha = 23^{\rm h} 30^{\rm m}$, $\delta = -55^{\circ}$ \\
  Field 2 (J2000) & $\alpha = 00^{\rm h} 30^{\rm m}$, $\delta = -55^{\circ}$ \\
  Area size for each field & $2^{\circ}\times 2^{\circ}$ \\
  Sensitivity (flux)& 0.18~mJy~beam$^{-1}$ \\
  Sensitivity (temperature)& 3.2~mK \\
  Gain & 17.3 K/Jy~beam$^{-1}$ \\
  \hline
  \end{tabular}
 \label{obsTab}
\end{table}
A standard procedure for the ATCA data reduction, as described in B03, was
followed to obtain Stokes $I$, $Q$, $U$, $V$ images.

Figure~\ref{dasi1_images} and \ref{dasi2_images} show the results for the
field 1 and 2, respectively. It is evident that the cleaning procedure did not
manage to remove completely the side lobes for the brightest sources even in
polarization.

The emission is dominated by point sources both in total intensity and
polarization. Besides this, no evident diffuse polarized emission rises up the
noise level in both fields. This is in contrast with that found in the B03 area
observed with the same array (same configuration and sensitivity). In order to
make this contrast evident, we plot the map of the Stokes parameter $U$ measured
in the B03 field for a direct comparison in Figure~\ref{dasi1_images} and
\ref{dasi2_images} (similar results hold for the $Q$ images). All the structures
present in the DASI fields seem to be associated to point-like sources and no
extended structures are evident like in the B03 field, where a polarized diffuse
emission of $\sim 11$~mK was detected spread throughout the area. Besides
intrinsic differences due to different sky positions, the lower emission found
there has been explained by the action of Faraday effects. These, by
transferring power from large to small angular scales, enhance the polarized
emission on the 3--30~arcmin scales the observations are sensitive to (Carretti
et al. 2005b).

The DASI fields, however, are at higher Galactic latitudes ($b\sim -58\degr$ and
$b\sim -62\degr$ for the two areas, respectively) and Carretti et al. (2005a)
found that the emission is unlikely to be affected by Faraday rotation effects
at 1.4~GHz at $|b| \ge 50^\circ$. This seems to be confirmed also by the recent
polarization map at 1.4~GHz of the Northern Celestial Hemisphere (Wolleben et
al. 2005).

We identify the point sources present in the two fields by setting a
detection threshold of 10$\sigma$ in order to avoid false and spurious
candidates due to the presence of residual non--white noise. The analysis
provides a catalogue complete down to a polarized flux $I_p = 2$~mJy (see
Table~\ref{source_positions_dasi1} and Table~\ref{source_positions_dasi2}).

We look for counterparts and find that every source, but the number six of
field 1, has a radio counterpart in the Parkes survey at 408~MHz (PKS,
Wright 1994), in the Parkes--MIT--NRAO survey at 4.85~GHz (PMN, Griffith \&
Wright 1993) or in the Sydney University Molonglo Sky Survey at 843~MHz (SUMSS,
Mauch et al. 2003). In the case of source number six we find a counterpart in
the optical/infrared Automated Plate Measuring survey (APM, Maddox et al. 1990).
When a source is present in more than one survey, the strongest one within
the beam area is listed.

Actually, we cannot justify the absence of the source number six in the
field 1 in the SUMSS survey. In fact, if we consider a power law behaviour for
the flux density $F_\nu \propto \nu^{\alpha_\nu}$ as a function of the frequency
$\nu$ and assume a spectral index $\alpha_\nu = -0.7$, this source would have a
flux of $\sim 195$~mJy at 843~MHz whereas the sensitivity of the SUMSS is 8~mJy.
Even in the case of a source fully self-absorbed, the spectral index would be
$\alpha_\nu = 2.5$ and the flux at 843~MHz would be $\sim 40$~mJy, at the limit
of a $5\sigma$ detection in the SUMSS. It is worth noting that the source is
unlikely to be in a complete self--absorbed regime because of its polarized
emission; then the spectral slope should be flatter.

Given the limited sample, no general conclusion can be drawn but just
a few qualitative remarks. We note that there are almost the same number of
sources in the two fields (18 and 15, respectively) at the flux limit $I_p =
2$~mJy and almost all the sources show a polarization degree of a few per cent
whereas only three sources are more than 10\% polarized. The mean polarization
percentage is 6.2\% for the first field and 3.8\% for the second one.
\begin{table*}
\begin{center}
\centering
\caption{Position, total intensity ($I$) and polarized intensity ($I^p$) of
	 the sources detected in the first DASI field.
	 The rms-error is the same for both the two intensities
	 and corresponds to the beam-sensitivity (0.18~mJy~beam$^{-1}$) but an
	 5\% error due to accuracy has to be added.
	 Polarization angle ($\phi$) and polarization degree ($\Pi = I^p/I$) 
         are also reported. The last column provides the distance from the
	 counterpart.}
\begin{tabular}{|c|c|c|c|c|c|c|c|c|l|c|}
source & RA & DEC & $I$ & $I^p$ & $\phi$ & $\Pi$ &
counterpart & distance\\
       & J2000 & J2000 & [mJy] & [mJy] &    &	 &    & [arcmin]\\
\hline
1	& $23^{\rm h} 22^{\rm m} 07^{\rm s}.1$ & $-54^\circ 45^\prime
29^{\prime\prime}.1$ & 1914.00 & 50.06 & $-33.9^\circ \pm 0.1^\circ$ & $2.6$\% &
PKS 2319-55 & 0.0\\
2	& $23^{\rm h} 33^{\rm m} 07^{\rm s}.4$ & $-54^\circ 16^\prime
12^{\prime\prime}.0$ & 296.10 & 20.76 & $-38.6^\circ \pm 0.3^\circ$ & $7.0$\% &
PKS 2330-545 & 0.1\\
3	& $23^{\rm h} 29^{\rm m} 34^{\rm s}.5$ & $-54^\circ 25^\prime
25^{\prime\prime}.2$ & 125.90 & 6.99 & $39.0^\circ \pm 0.7^\circ$  & $5.6$\% &
SUMSS J232942-542524 & 1.2\\
4	& $23^{\rm h} 28^{\rm m} 04^{\rm s}.8$ & $-55^\circ 41^\prime
11^{\prime\prime}.2$ & 99.84 & 6.73 & $11.9^\circ \pm 0.8^\circ $ & $2.4$\% &
SUMSS J232806-554110 & 0.2\\ 
5	& $23^{\rm h} 28^{\rm m} 35^{\rm s}.4$ & $-54^\circ 41^\prime
15^{\prime\prime}.0$ & 55.91 & 6.30 & $-27.6^\circ \pm 0.8^\circ$  & $11.27$\% &
SUMSS J232835-544125 & 0.2\\
6	& $23^{\rm h} 27^{\rm m} 27^{\rm s}.2$ & $-54^\circ 42^\prime
31^{\prime\prime}.3$ &  136.70 &  5.50 & $-2.4^\circ \pm 0.9^\circ$ & $4.0$\% &
APMUKS(BJ) B232445.84-545734.8 & 1.7\\
7	& $23^{\rm h} 33^{\rm m} 29^{\rm s}.3$ & $-54^\circ 09^\prime
25^{\prime\prime}.4$ &  57.30 &  4.76 & $-10.2^\circ \pm 1.1^\circ$ & $8.3$\% &
SUMSS J233329-540934 & 0.1\\
8	& $23^{\rm h} 27^{\rm m} 24^{\rm s}.8$ & $-54^\circ 51^\prime
01^{\prime\prime}.0$ & 105.60 &  4.45 & $-26.5^\circ \pm 1.2^\circ$ & $4.2$\% &
SUMSS J232724-545114 & 0.2\\
9	& $23^{\rm h} 29^{\rm m} 25^{\rm s}.4$ & $-54^\circ 54^\prime
22^{\prime\prime}.5$ &  209.00 &  4.31 & $-8.7^\circ \pm 1.2^\circ$ & $2.1$\% &
SUMSS J232925-545434 & 0.2\\
10	& $23^{\rm h} 32^{\rm m} 07^{\rm s}.6$ & $-54^\circ 44^\prime
06^{\prime\prime}.1$ &  63.00 &  4.27 & $35.0^\circ \pm 1.2^\circ$ & $6.8$\% &
SUMSS J233207-544406 & 0.1\\
11	& $23^{\rm h} 26^{\rm m} 14^{\rm s}.4$ & $-54^\circ 03^\prime
07^{\prime\prime}.7$ &  57.60 &  3.70 & $-22.3^\circ \pm 1.4^\circ$ & $6.4$\% &
SUMSS J232614-540321 & 0.2\\
12	& $23^{\rm h} 30^{\rm m} 33^{\rm s}.1$ & $-54^\circ 32^\prime
06^{\prime\prime}.4$ &  24.02 &  3.63 & $14.3^\circ \pm 1.4^\circ$ & $15.0$\% &
SUMSS J233033-543141 & 0.4\\
13	& $23^{\rm h} 34^{\rm m} 44^{\rm s}.3$ & $-54^\circ 19^\prime
00^{\prime\prime}.7$ &  43.37 &  3.32 & $34.2^\circ \pm 1.6^\circ$ & $7.7$\% &
SUMSS J233445-541910 & 0.2\\
14	& $23^{\rm h} 27^{\rm m} 20^{\rm s}.3$ & $-55^\circ 09^\prime
16^{\prime\prime}.9$ &  104.20 &  2.86 & $-31.6^\circ \pm 1.8^\circ$ & $2.7$\% &
SUMSS J232718-550936 & 0.5\\
15	& $23^{\rm h} 31^{\rm m} 20^{\rm s}.1$ & $-55^\circ 27^\prime
06^{\prime\prime}.4$ &  12.84 &  2.67 & $-26.3^\circ \pm 1.9^\circ$ & $20.8$\% &
SUMSS J233119-552702 & 0.1\\
16	& $23^{\rm h} 32^{\rm m} 53^{\rm s}.9$ & $-55^\circ 11^\prime
16^{\prime\prime}.2$ &  106.10 &  2.31 & $-34.5^\circ \pm 2.2^\circ$ & $2.2$\% &
SUMSS J233253-551055 & 0.4\\
17	& $23^{\rm h} 25^{\rm m} 20^{\rm s}.1$ & $-55^\circ 02^\prime
17^{\prime\prime}.5$ &  148.20 &  2.21 & $6.5^\circ \pm 2.3^\circ$ & $1.5$\% &
SUMSS J232520-550228 & 0.2\\
18	& $23^{\rm h} 34^{\rm m} 05^{\rm s}.1$ & $-54^\circ 11^\prime
00^{\prime\prime}.6$ &  131.20 &  2.13 & $25.3^\circ \pm 2.4^\circ$ & $1.6$\% &
SUMSS J233404-541058 & 0.0\\
\hline
\label{source_positions_dasi1}
\end{tabular}
\end{center}
\end{table*}
\begin{table*}
\begin{center}
\centering
\caption{As for Table~\ref{source_positions_dasi1} but for field 2}
\begin{tabular}{|c|c|c|c|c|c|c|l|c|}
source & RA & DEC & $I$ & $I^p$ & $\phi$ & $\Pi$ &
counterpart & distance\\
       & J2000 & J2000 & [mJy] & [mJy] &    &	 &    & [arcmin]\\
\hline
1	& $00^{\rm h} 31^{\rm m} 02^{\rm s}.8$ & $-55^\circ 21^\prime
12^{\prime\prime}.5$ & 264.10 & 15.66 & $26.0^\circ \pm 0.3^\circ$ & $5.9$\% &
PKS 0028-556 & 0.1\\
2	& $00^{\rm h} 30^{\rm m} 33^{\rm s}.6$ & $-54^\circ 38^\prime
01^{\prime\prime}.2$ & 608.50 & 15.54 & $-9.3^\circ \pm 0.3^\circ$  & $2.6$\% &
PKS 0028-549 & 0.2\\
3	& $00^{\rm h} 32^{\rm m} 26^{\rm s}.3$ & $-54^\circ 54^\prime
55^{\prime\prime}.3$ & 493.00 & 11.71 & $6.1^\circ \pm 0.4^\circ$  & $2.4$\% &
PKS 0030-552 & 0.2\\
4	& $00^{\rm h} 36^{\rm m} 54^{\rm s}.0$ & $-55^\circ 36^\prime
28^{\prime\prime}.9$ & 196.90 & 11.25 & $15.3^\circ \pm 0.5^\circ $ & $5.7$\% &
SUMSS J003653-553632 & 0.1\\
5	& $00^{\rm h} 26^{\rm m} 14^{\rm s}.7$ & $-55^\circ 23^\prime
38^{\prime\prime}.4$ & 105.50 & 8.51 & $-21.3^\circ \pm 0.6^\circ$ & $8.1$\% &
SUMSS J002615-552337 & 0.1\\
6	& $00^{\rm h} 35^{\rm m} 57^{\rm s}.3$ & $-54^\circ 03^\prime
23^{\prime\prime}.0$ &  177.40 &  8.27 & $-4.7^\circ \pm 0.6^\circ$ & $4.7$\% &
SUMSS J003558-540324 & 0.1\\
7	& $00^{\rm h} 32^{\rm m} 55^{\rm s}.3$ & $-54^\circ 56^\prime
28^{\prime\prime}.0$ &  1381.00 &  7.09 & $7.6^\circ \pm 0.7^\circ$ & $0.5$\% &
SUMSS J003254-545614 & 0.3\\
8	& $00^{\rm h} 33^{\rm m} 20^{\rm s}.3$ & $-55^\circ 37^\prime
54^{\prime\prime}.9$ & 134.10 &  6.61 & $-39.6^\circ \pm 0.8^\circ$ & $4.9$\% &
SUMSS J003319-553757 & 0.1\\
9       & $00^{\rm h} 23^{\rm m} 36^{\rm s}.9$ & $-54^\circ 31^\prime
13^{\prime\prime}.2$ &  98.59 &  6.26 & $-15.8^\circ \pm 0.8^\circ$ & $6.3$\% &
SUMSS J002340-543151 & 0.9\\
10	& $00^{\rm h} 22^{\rm m} 40^{\rm s}.9$ & $-55^\circ 03^\prime
58^{\prime\prime}.4$ &  120.60 &  4.51 & $-23.3^\circ \pm 1.1^\circ$ & $3.7$\% &
SUMSS J002241-550400 & 0.0\\
11	& $00^{\rm h} 35^{\rm m} 53^{\rm s}.1$ & $-54^\circ 16^\prime
47^{\prime\prime}.1$ &  146.50 &  4.31 & $28.7^\circ \pm 1.2^\circ$ & $2.9$\% &
SUMSS J003554-541634 & 0.3\\
12	& $00^{\rm h} 25^{\rm m} 10^{\rm s}.7$ & $-54^\circ 27^\prime
39^{\prime\prime}.9$ &  131.40 &  3.56 & $-39.1^\circ \pm 1.5^\circ$ & $2.7$\% &
PMN J0025-5427 & 0.0\\
13	& $00^{\rm h} 28^{\rm m} 15^{\rm s}.2$ & $-55^\circ 26^\prime
43^{\prime\prime}.6$ &  138.70 &  3.36 & $-40.8^\circ \pm 1.5^\circ$ & $2.4$\% &
SUMSS J002814-552701 & 0.3\\
14	& $00^{\rm h} 23^{\rm m} 03^{\rm s}.4$ & $-55^\circ 42^\prime
48^{\prime\prime}.3$ &  159.80 &  3.32 & $8.4^\circ \pm 1.6^\circ$ & $2.1$\% &
PMN J0023-5542 & 0.3\\
15	& $00^{\rm h} 33^{\rm m} 44^{\rm s}.8$ & $-54^\circ 56^\prime
24^{\prime\prime}.8$ &  152.70 &  3.23 & $-37.1^\circ \pm 1.6^\circ$ & $2.1$\% &
SUMSS J003344-545619 & 0.1\\
\hline
\label{source_positions_dasi2}
\end{tabular}
\end{center}
\end{table*}

\section{Power spectrum analysis}
\label{CMBsec}
We compute the power spectra of the $E$-- and $B$--modes in the
inner $2^\circ \times 2^\circ$ square of both fields, where the sensitivity is
higher.
\begin{figure*}
  \includegraphics[angle=0, width=0.4\hsize]{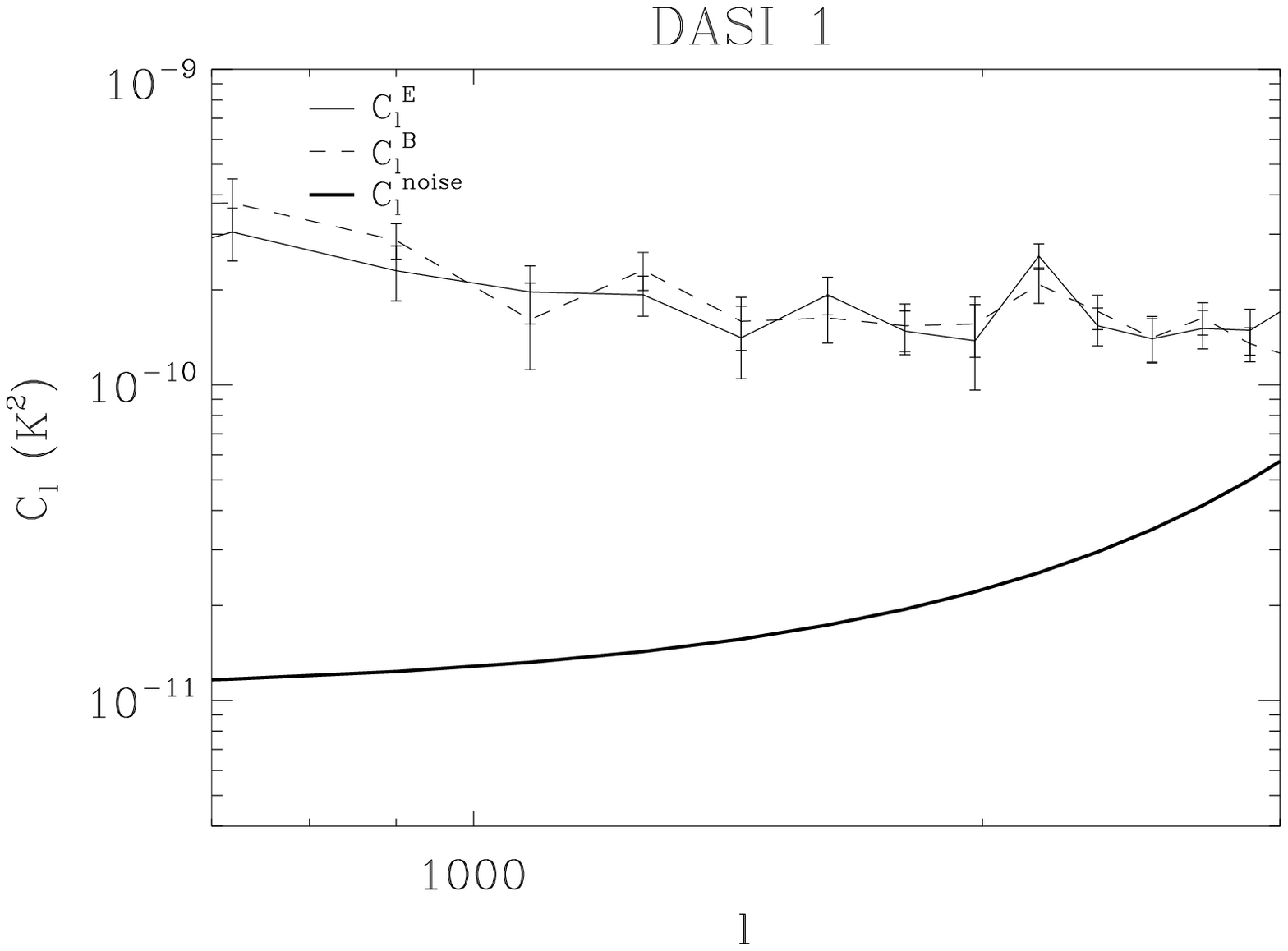}
  \includegraphics[angle=0, width=0.4\hsize]{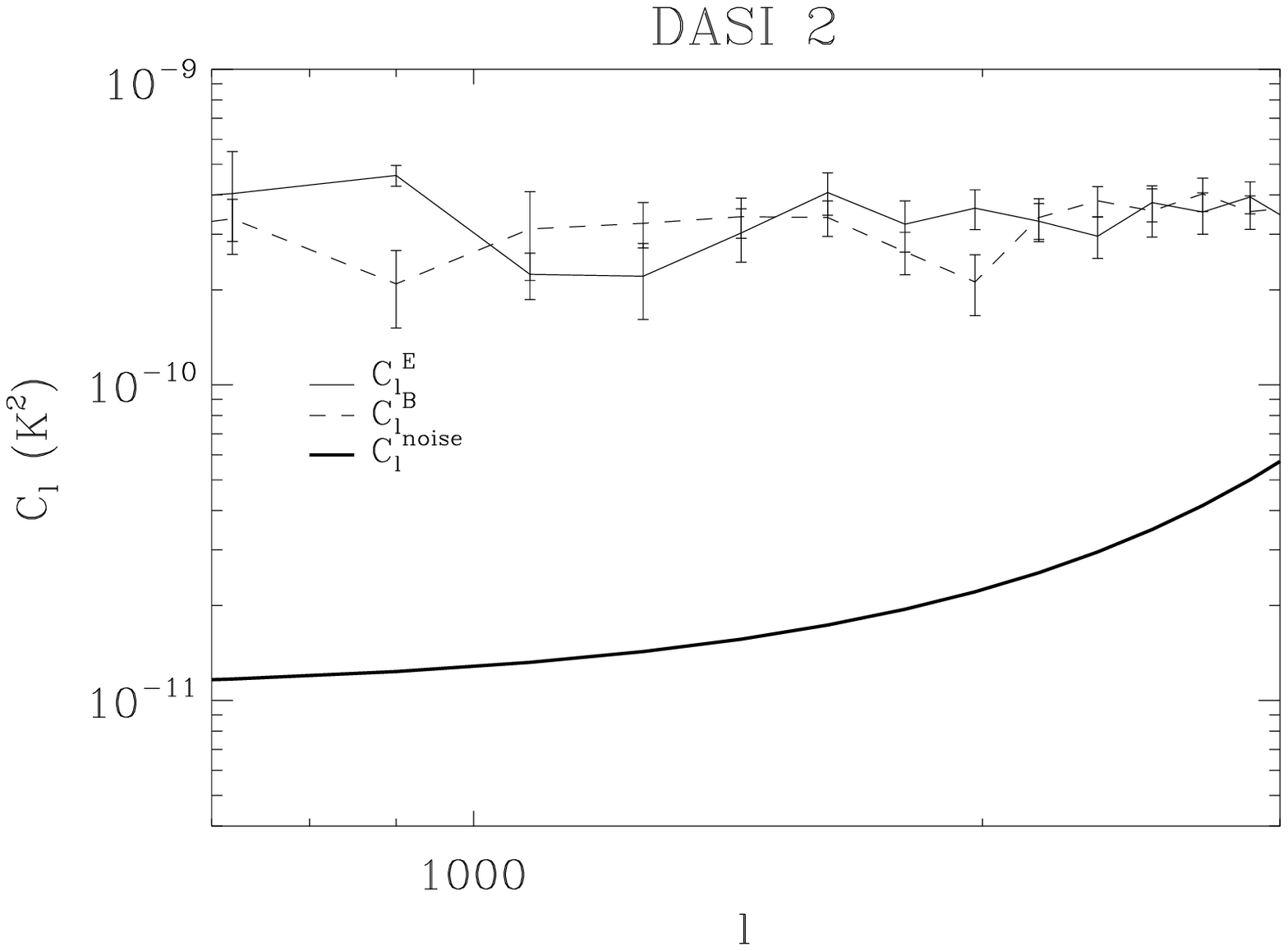}
\caption{Power spectra of field 1 (left) and field 2 (right) with the respective
$1\sigma$ error bars. Solid line represet the noise (see
text).\label{power_spectra_PS}}
\end{figure*}
\begin{figure*}
  \includegraphics[angle=0, width=0.4\hsize]{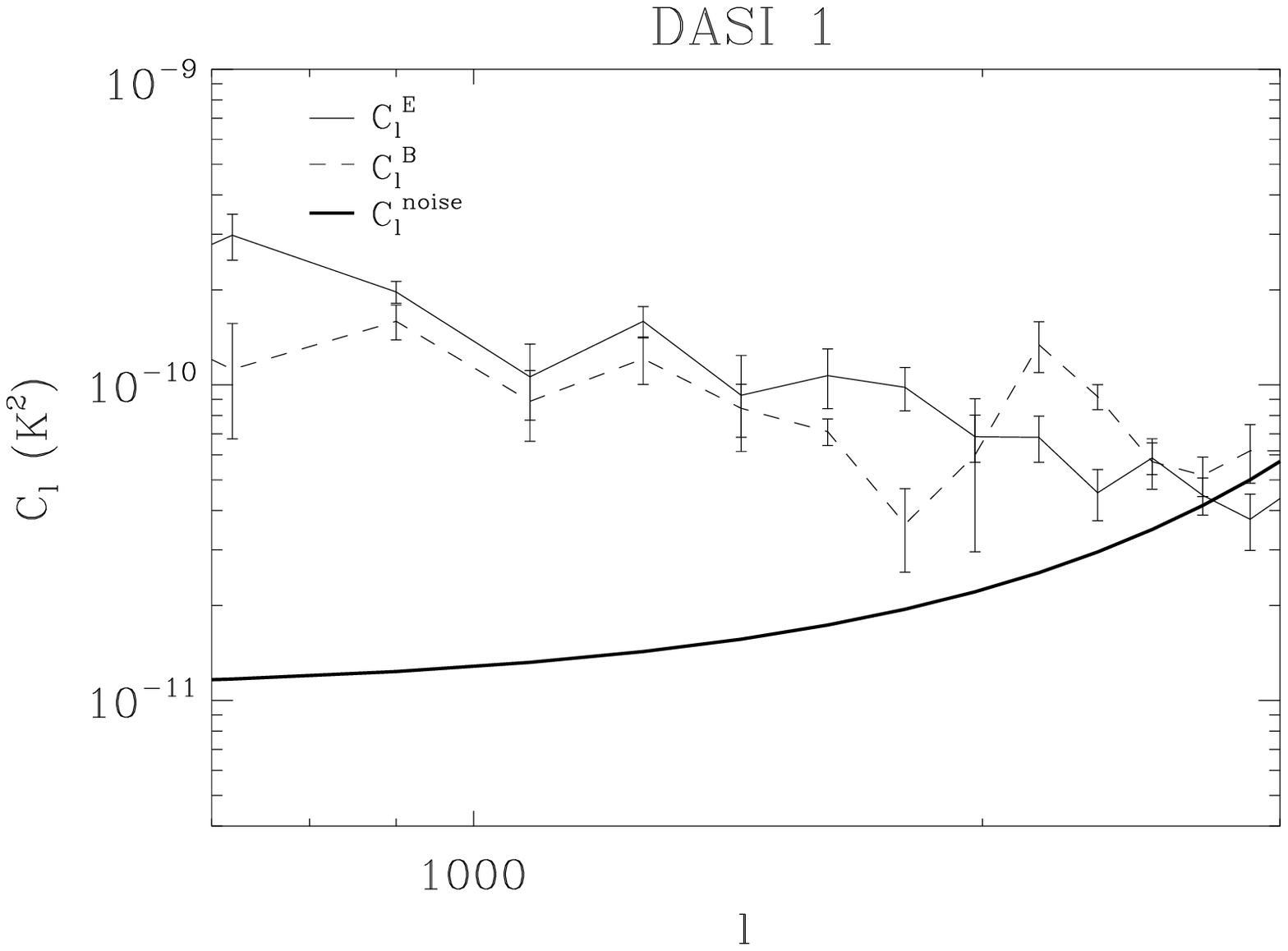}
  \includegraphics[angle=0, width=0.4\hsize]{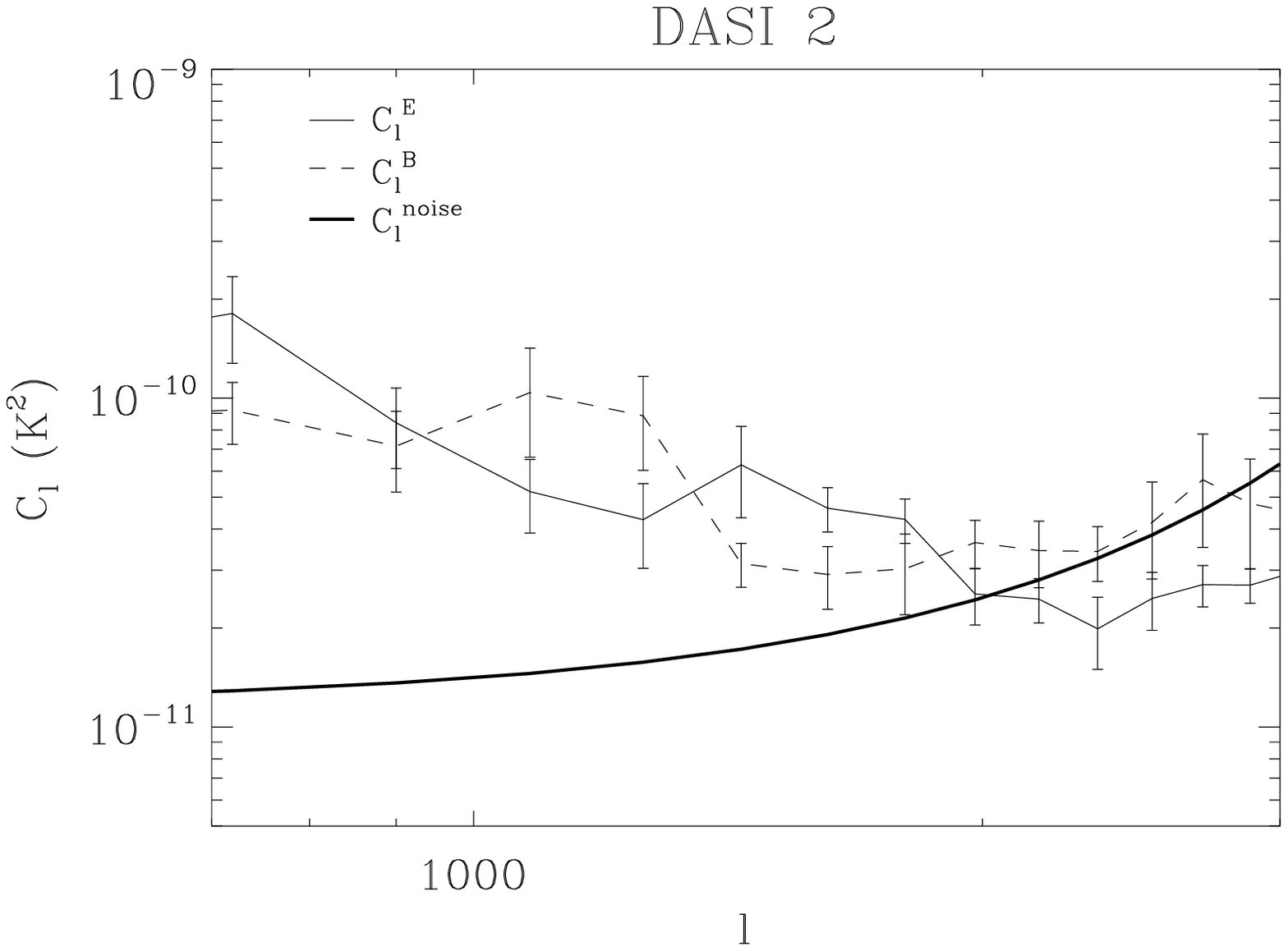}
\caption{As for Figure\ref{power_spectra_PS} but after point sources
subtraction.\label{power_spectra}}
\end{figure*}
The power spectra are plotted in Figure~\ref{power_spectra_PS} together with the
noise contribution defined by (Tegmark 1997) 
\begin{equation}
    C_{\ell}^{noise} = \frac{f_{sky} \, 4 \pi \sigma^2_P}{N \, B_\ell^2}
\end{equation}
where $f_{sky}$ is the sky fraction, $N$ is the number of pixels, $B_\ell^2$ is
the power spectrum of the beam--window function and $\sigma_P$ the pixel
sensitivity.

The power spectra fit the power law equation
\begin{equation}
    C_{\ell}^X = C_{2000}^{X} \left({\ell \over 2000}\right)^{\beta_X},
    {\;\;} X=E,\,B
\end{equation}
as a function of the multipole $\ell$. Best fit parameters are shown in
Table~\ref{powFitTab_PS}. The slopes of the spectra for the field 2 are
compatible with a flat $\beta = 0$ spectrum within $2\sigma$ C.L., which is
typical of the emission dominated by point sources. A similar result holds for
the $E$--mode of the field 1, whereas the $B$--mode looks steeper and compatible
with a point source power spectrum only at $\sim 3\sigma$ C.L. Also the spectra,
therefore, support an emission dominated by point sources.

To estimate an upper limit of the diffuse emission contribution, we subtracted
all the bright sources listed in Table~\ref{source_positions_dasi1}
and \ref{source_positions_dasi2}. The resulting power spectra are shown in
Figure~\ref{power_spectra}.

It can be seen a remarkable lowering of the flux and a certain steepening of the
slopes, in particular for field 2. This is clearly shown by the best fit
parameters quoted in Table~\ref{powFitTab}.
\begin{table}
 \centering
  \caption{Fit parameters for $E$ and $B$ spectra plotted in
  Figure~\ref{power_spectra_PS}. The fit has been performed in 
  the $800 < \ell < 2800$ multipole range.\label{specTab}}
  \begin{tabular}{@{}lccc@{}}
  \hline
  Spectrum & \# Field & $C_{2000}^X$~($\mu$K$^2$) & $\beta_X$ \\
  \hline
  $C^E_\ell$ &  1  &  $183 \pm 9\phantom{0}$ & $-0.19 \pm 0.15$ \\
  $C^B_\ell$ &  1  &  $177 \pm 9\phantom{0}$ & $-0.44 \pm 0.13$ \\
  $C^E_\ell$ &  2  &  $345 \pm 17$ & $-0.20 \pm 0.10$ \\
  $C^B_\ell$ &  2  &  $340 \pm 16$ & $+0.34 \pm 0.16$ \\
  \hline
  \end{tabular}
 \label{powFitTab_PS}
\end{table}
\begin{table}
 \centering
  \caption{As for Table~\ref{powFitTab_PS} but after point
  sources subtraction.\label{specTab}}
  \begin{tabular}{@{}lccc@{}}
  \hline
  Spectrum & \# Field & $C_{2000}^X$~($\mu$K$^2$) & $\beta_X$ \\
  \hline
  $C^E_\ell$ &  1  &  $ 73 \pm 4$ & $-1.29 \pm 0.10$ \\
  $C^B_\ell$ &  1  &  $ 80 \pm 4$ & $-0.63 \pm 0.13$ \\
  $C^E_\ell$ &  2  &  $ 32 \pm 2$ & $-1.09 \pm 0.20$ \\
  $C^B_\ell$ &  2  &  $ 36 \pm 3$ & $-0.58 \pm 0.25$ \\
  \hline
  \end{tabular}
 \label{powFitTab}
\end{table}

Although these spectra certainly exhibit a lower contribution by point sources
they cannot be considered as representative of the diffuse emission. In fact,
the spectra match the expected noise level at the high $\ell$--range tail,
so that the power on the largest scales can likely be due to residual
non--white noise. As a consequence, we prefer to consider these spectra as upper
limits of the diffuse emission.

\section{Discussion}
\label{discussion_CMB}
\begin{figure}
  \includegraphics[angle=0, width=1\hsize]{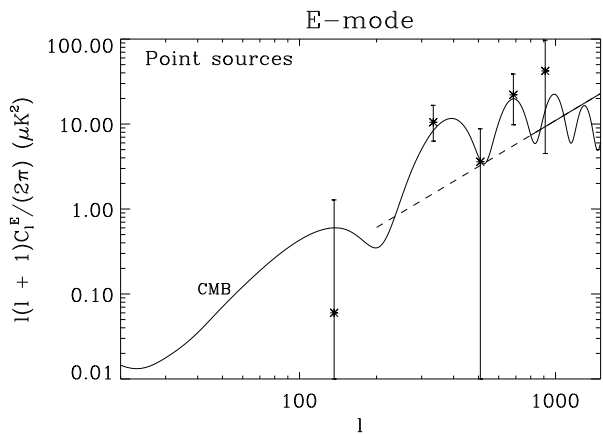}
\caption{The CMB $E$--mode power spectrum (solid line) computed with the WMAP
first year cosmological parameters (Spergel et al. 2003), the measurements
reported by the DASI team (asterisks, Leitch et al. 2005), the extrapolation to 
30~GHz of the measured power spectra in the second DASI field (solid stright
line) and its extention down to $\sim 1^\circ$ angular scale (dashed
line).\label{cmb_synch_power_spectra_PS}}
\end{figure}

The spectra reported in Figure~\ref{power_spectra_PS} allow an estimate of the
contamination to the CMB due to point source emission. Then, we use the
$E$--mode power spectrum (Figure~\ref{power_spectra_PS}) of the field 2, that,
having the highest signal, represents the worst case. The frequency
extrapolation is not trivial because of the large range of possible spectral
slopes. In low frequency surveys, the radio 
population is a mixture of steep ($\alpha_\nu < -0.5$) and flat
spectrum ($\alpha_\nu > -0.5$) sources. In particular, the steep spectrum
sources have been found for the 87\% of the population in the NRAO VLA Sky
Survey (NVSS, Condon et al. 1998), whereas the remaining 13\% are flat spectrum
sources. Therefore, we take the 13\% of the $E$--mode power spectrum and 
scale it up to 30~GHz with a brightness temperature spectral index $\alpha =
-2.25$, which is an average value in the interval $-2.5 < \alpha < -2.0$ quoted by
Toffolatti et al. (1998) for flat--spectrum sources. Then, we take the
other 87\% of the $E$--mode power spectrum and scale it up with $\alpha =
-2.75$, as reported by Peacock \& Gull 
(1981) for steep spectrum sources. The two resulting contributions are added
together to give the estimate of the $E$--mode power spectrum at 30~GHz. Since
the angular behaviour is compatible with a point source power spectrum, we
extend it down to $\ell = 200$. The result is shown in 
Figure~\ref{cmb_synch_power_spectra_PS} together with the DASI measurements and
the CMBP $E$-mode power spectrum expected according to cosmological parameters
measured after WMAP data (Spergel et al. 2003).

The situation represented in Figure~\ref{cmb_synch_power_spectra_PS} does not
exclude a point source contamination to the CMB $E$--mode power spectrum. At
$\ell = 300$ we find $\ell(\ell + 1)C^E_\ell/2\pi \simeq 1.3$~$\mu$K$^2$, which
is just slightly greater than the value quoted as upper limit by Leitch 
et al. (2005, 0.98~$\mu$K$^2$ at $2\sigma$ C.L.). At higher multipoles, the
estimated point source emission is increasing and, compared to the CMB 
peaks, ranges from a factor 10 to a factor 3 lower than the CMB spectrum (3.2
and 1.7 in signal, respectively). Although within the limits of the
uncertainties in the frequency extrapolation, our results suggest that the point
source contamination in the DASI data is not neglibile. In the light of this, a
re--analysis of the data considering a subtraction of the detected sources could
allow a better estimate of the extragalactic signal component.

\begin{figure}
  \includegraphics[angle=0, width=1\hsize]{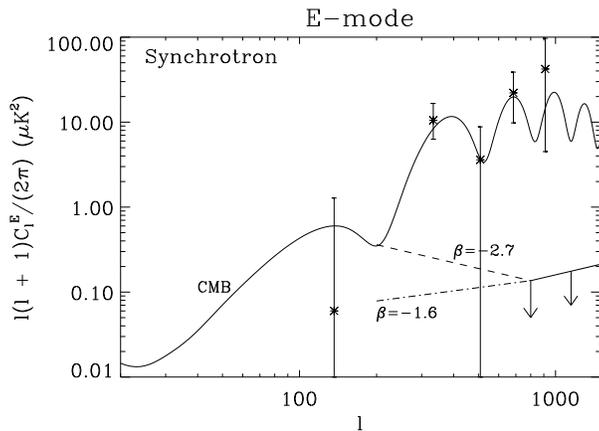}
\caption{As Figure~\ref{cmb_synch_power_spectra_PS} but for the diffuse
synchrotron emission: the extrapolation to 
30~GHz of the upper limit on the synchrotron power spectrum obtained by our
1.4~GHz observations (solid straight line) is plotted with two possible
extrapolations of the synchrotron power spectrum up to $\sim 1^\circ$ angular
scale with a slope $\beta = -2.7$ (dashed line) and a slope $\beta = -1.6$ (dotted
dashed line, see text for details). The arrows indicates that our extrapolation
represents an upper limit.\label{cmb_synch_power_spectra}}
\end{figure}

Power spectra of Figure~\ref{power_spectra} allow us to estimate an upper
limit on the diffuse synchrotron emission as contaminant of the CMBP. We use the
result of the field 1, which represents the worst case. The power spectrum of
field 1 is scaled up to
30~GHz using a brightness temperature spectral index $\alpha = -3.1$ (Bernardi
et al. 2004), and the result is shown in Figure~\ref{cmb_synch_power_spectra}.
Since the multipole range accessible by our and DASI data are only marginally
overlapped, we perform an extrapolation in the $\ell$-space. Since the spectrum
we measure is an upper limit and prevents us to use its slope for such
extrapolation, we consider a pessimistic and a more realistic case.

We use a slope $\beta = -2.7$, that is the steepest slope measured so far for
the synchrotron $E$-mode (Tucci et al. 2002), as a worst case. This value
measured on the Galactic plane at 1.4~GHz is likely to be significantly 
alterated by Faraday effects. However, at the latitudes of the DASI fields, a
value $\beta = -1.6 \pm 0.2$ is more likely expected (Bruscoli et al. 2002),
which is more typical where Faraday rotation effects are weak. Therefore, we
perform a second extrapolation using $\beta = -1.6$.

The results (Figure~\ref{cmb_synch_power_spectra}) show that the expected
theoretical CMBP $E$-mode power spectrum is not contaminated down to the peak at
$\ell \sim 300$. Considering the most likely slope of $\beta = -1.6$ we see that
neither the theoretical peak at $\ell \sim 120$ would not be contamined by the
synchrotron emission. The average value of the synchrotron power spectrum over
the $200 < \ell < 1050$ multipole range is $\sim 0.19$~$\mu$K$^2$, which is a
factor $\sim 5$ lower than the upper limit quoted by the DASI team
(0.91~$\mu$K$^2$, see Leitch et al. 2004). This furtherly supports the
evaluation of negligible synchrotron contamination done by the DASI team itself.

In the light of the data presented in this paper, a twofold picture is emerging.
Even if we did not detect the synchrotron diffuse emission at 1.4~GHz, the upper
limit we set is robust enough to conclude that this component is not a serious
contaminant of the CMB $E$--mode at 30~GHz. For this reason, our observations
support the conclusions inferred by Kovac et al. (2002) and Leitch et al.
(2005).

On the other hand, we argue that the problem regarding the point
source contamination is still open. Leitch et al. (2005) find an upper limit of
0.98~$\mu$K$^2$ and exclude a contamination by point sources. From the data
presented here, we retrieve a value very close to it but we cannot exclude the
CMBP power spectrum is not contaminated at $\ell > 300$
(Figure~\ref{cmb_synch_power_spectra_PS}). Our data suggest the CMBP
signal should dominate over the point source emission at 30~GHz but a
reappraisal of the their contamination is recommended.

\section*{Acknowledgments}

We thank Marijke Haverkorn for useful discussions and suggestions. This work has
been carried out in the framework of the SPOrt experiment, a programme funded by
ASI (Italian Space Agency). The Australia Telescope Compact Array is part of the
Australia Telescope, which is funded by the Commonwealth of Australia for
operation as a National Facility managed by CSIRO. We acknowledge the use of the
CMBFAST package. This research has made use of the NASA/IPAC Extragalactic
Database (NED) which is operated by the Jet Propulsion Laboratory, California
Institute of Technology, under contract with the National Aeronautics and Space
Administration.

\bsp

\label{lastpage}

\end{document}